\documentstyle [12pt] {article}
\title{SPONTANEOUS BREAKING OF THE QUANTUM SUPERPOSITION }
\author{Vladan Pankovi\'c, Milan Predojevi\'c \\
Department of Physics, Faculty of Sciences \\21000 Novi Sad, Trg Dositeja Obradovi\'ca 4, Serbia \\
vdpan@neobee.net}

\date {}
\begin{document}
\maketitle

\vspace {0.3cm}

PACS number: 03.65.Ta

\vspace{0.5cm}

\begin {abstract}

In this work spontaneous (non-dynamical) breaking (effective
hiding) of the unitary quantum mechanical dynamical symmetry
(superposition) is considered. It represents an especial but very
interesting case of the general formalism of the spontaneous
symmetry breaking (effective hiding). Conceptual analogies with
spontaneous breaking of the gauge symmetry in Weinberg-Sallam's
electro-weak interaction are pointed out. Also, consequences of
the spontaneous superposition breaking in the measurement process
are discussed. Measurement can be considered as a typical Landau's
continuous phase transition with spontaneous superposition
breaking (effective hiding) effectively corresponding to collapse.
Here critical values of the order parameters are determined by
Heisenberg's uncertainty relations.
\end {abstract}
\vspace {0.5cm}

{\it   "Nothing is sure for me but what's uncertain:

        Obscure, whatsever is plainly clear to see:

         I've no doubt, except of everything certain:

         Science is what happens accidentally:

         I winn it all, yet a loser I'm bound to be:"

\vspace{0.5cm}

Francois Villon (1431. - 1463?),

Ballade: Du Concurs De Blois}

\vspace {0.5cm}

\section {Introduction}

As it is well-known, any dynamical symmetry transformation
determines a reference system, i.e. referential frame for
description of the dynamical processes. For this reason unbroken
dynamical symmetry expresses the equivalence (relativity) of all
referential frames, while breaking of a dynamical symmetry
corresponds to preference (absoluteness) of some referential
frame.

In the physics there are two principally different kinds of the
breaking of a dynamical symmetry. First one represents the
dynamical breaking and second one spontaneous (non-dynamical)
breaking (effective hiding) [1]-[3].

Dynamical breaking of a dynamical symmetry represents, roughly
speaking, the breaking of the dynamical symmetry by hidden
dynamical terms. Precisely, by dynamical breaking of a dynamical
symmetry  there is one exact dynamics and an its approximation,
i.e. approximate dynamics.

Exact dynamics yields an exact, i.e. with zero absolute error,
deterministic description of corresponding physical process. Also,
symmetries of the exact dynamics stand completely conserved during
time.

Approximate dynamics represents, mostly, an approximate Taylor
expansion of the exact dynamics statistically averaged over some
terms. (For this reason given terms become effectively hidden in
given approximate dynamics.) Consistent approximate dynamics needs
that corresponding Taylor expansion is convergent globally, i.e.
in whole space of the basic dynamical variables. Also, an interval
statistical estimation [4] of the exact dynamics needs not only an
average value, i.e. consistent approximate dynamics. It needs,
also, a standard deviation (square root of the dispersion) that
can be considered as as the absolute error of the approximate
dynamics.

Consider a case when consistent approximate dynamics, i.e.
convergent average value is sufficiently larger than standard
deviation, i.e. absolute error of the approximate dynamics. Then
it can be stated that approximate dynamics is self-complete. It
means that, at the approximate level of the analysis, only
approximate dynamics is sufficient for a complete, deterministic
description of the physical process. Namely, here corresponding
absolute error, that implies necessity of the statistical
description, is neglectable. (I.e. here statistical interval
estimation can be  approximately consistently reduced in the point
statistical estimation.) In the same case symmetries of the
approximate dynamics stand conserved during time.

Consider now the opposite case when consistent approximate
dynamics, i.e.  convergent average value becomes proportional or
smaller than standard deviation, i.e. absolute error. Then
approximate dynamics becomes self-incomplete. It means that
approximate dynamics is not sufficient for complete approximate
description of the physical process. Namely, here corresponding
absolute error, that implies necessity of the statistical
description, is not neglectable. It implies that some of the
symmetries of the approximate dynamics are unsatisfied for
self-incomplete approximate dynamics.

It is possible that a self-complete approximate dynamics turns
during time in a self-incomplete approximate dynamics. Namely,
suppose that initially there is unique approximate dynamical
state, i.e. dynamical state of the self-complete approximate
dynamics. Also, suppose that this state represents a statistical
mixture of the exact dynamical states, i.e. dynamical states of
the exact dynamics. Given statistical mixture exactly dynamically
evolves according to exact dynamics. It corresponds to approximate
dynamical evolution of the approximate dynamical state according
to approximate dynamics. Suppose that in a critical time moment
self-complete approximate dynamics turns in the self-incomplete
approximate dynamics. Precisely, suppose that then unique
approximate dynamical state turns in a statistical mixture of the
approximate dynamical states. Roughly speaking, initially hidden,
i.e. implicit statistical mixture of the (exact) dynamical states
becomes finally, i.e. in given critical time moment explicit (at
the approximate level of the analysis). It corresponds to
realization of the probabilistic events. Also, it corresponds to a
typical Landau's continuous phase transition [5]. In given phase
transition critical values of the order parameters correspond to
given critical time moment. Of course given phase transition
includes a breaking of the symmetries of the self-complete
approximate dynamics. Given breaking is caused by exact dynamical
terms. Given terms are hidden at the approximate level of the
analysis corresponding to approximate dynamics. For this reason
given symmetry breaking is called dynamical symmetry breaking.

Dynamical breaking of the dynamical symmetry exists, for example,
by parity breaking in the weak (nuclear) interaction [2] etc.
Also, dynamical breaking of the dynamical symmetries exists in the
simple classical physical models of the different gambling games,
eg. tossing of a coin or playing at dice etc. Generally speaking,
dynamical breaking of the dynamical symmetries represents one of
the simplest physical model of the mathematical theory of the
probability (statistics) [4].

Spontaneous breaking of a dynamical symmetry represents, roughly
speaking, choice of an asymmetric solution  of the dynamically
symmetric equation. But it needs a more detailed explanation. By
spontaneous breaking of a dynamical symmetry, also, there is one
exact dynamics and an approximate dynamics. In any time moment
exact dynamics is exactly satisfied and its symmetries are always
exactly conserved.

Approximate dynamics, again, represents, mostly, an approximate
Taylor expansion of the exact dynamics statistically averaged over
some terms.  Also, a standard deviation representing the absolute
error of the approximate dynamics can be introduced. Meanwhile,
approximate dynamics is here globally inconsistent or globally
unstable. It means the following. At least for some values of the
basic dynamical variables, or, at least in some domains in the
space, approximate Taylor expansion of the exact dynamics
diverges. However, approximate dynamics can be locally consistent,
i.e. locally stable. It means that for some values of the basic
dynamical variables or in some, local, domains of the space
approximate Taylor expansion of the approximate dynamics
converges. But, some principal characteristic, eg. symmetries of
the exact dynamics can  forbid split of the approximate dynamical
state in its simultaneously existing locally stable parts, i.e.
locally stable approximate dynamical states. Nevertheless, an
additional probabilistic, i.e. statistical approximation admits
that a locally stable approximate dynamical state be effectively
treated as a globally stable approximate dynamical state. Or,
generally, globally unstable approximate dynamical state can be,
by additional statistical approximation, i.e. with corresponding
probability, effectively changed by, i.e. presented or projected
in some globally stable approximate dynamical state. Last state
corresponds to some locally stable approximate dynamical state
too. In this way globally unstable approximate dynamical state and
statistically obtained effective globally stable dynamical state
exist simultaneously, but at the discretely different approximate
levels of the analysis accuracy. First level does not admit a
consistent approximate dynamical description of the physical
system. Other level, whose definition includes some statistical
methods, admits, effectively, consistent approximate dynamical
description of the physical system. For this reason there is none
dynamical transition from globally unstable in some effectively
globally stable dynamical state. Realization of the probabilistic
events corresponds simply to the change of the first by second
approximate level of the analysis accuracy. In this way different
effectively globally stable approximate dynamical states that
appear with corresponding probabilities correspond implicitly to
the same globally unstable approximate state or to exact dynamical
state.

Suppose that an exact dynamical state initially can be
approximated by a globally stable approximate dynamical state.
Suppose that this exact dynamical state exactly dynamically
evolves during time. It corresponds to approximate dynamical
evolution of globally stable approximate dynamical state. Suppose
that in a critical time moment exact dynamical state, in
corresponding approximation, becomes a globally unstable
approximate dynamical state. Further consistent approximate
dynamical description of the exact dynamical state needs, in the
practically same critical time moment, mentioned change of the
first by the second level of the analysis accuracy. Or, it needs
mentioned non-dynamical change of the globally unstable
approximate dynamical state by one statistically obtained
effectively globally stable dynamical state. Complex change of the
initial globally stable approximate dynamical state by final
effectively globally stable approximate dynamical state includes
both approximate dynamical evolution and change of the approximate
levels of the analysis accuracy. Such change corresponds to a
typical Landau's continuous phase transition [5]. In given phase
transition critical values of the order parameters correspond to
given critical time moment. Of course given phase transition
includes a breaking of the symmetries of the approximate dynamics.
Given breaking is non-dynamical, i.e. it is not caused by exact
dynamical terms hidden within given approximation. It is
practically caused by a non-dynamical change of the first by the
second approximate level of the analysis accuracy. For this reason
given non-dynamical symmetry breaking is called spontaneous
symmetry breaking, i.e. effective hiding. It means that here
approximate dynamical symmetry is not really broken at the first
level of the approximate analysis accuracy. It is only effectively
hidden by statistical change of the first by the second level of
the approximate analysis accuracy.

Spontaneous breaking (effective hiding)  of the dynamical symmetry
exists, for example, by Heisenberg's ferromagnetics, by
Weinberg-Sallam's electro-weak interaction, etc. [1]-[3]. (There
is wide-spread phrase that Higgs boson breaks gauge symmetry in
the electro-weak interaction. It, seemingly, implies that gauge
symmetry is dynamically broken by Higgs boson. Meanwhile, as it is
well-known [1]-[3], given phrase represents a simplification only.
Appearance of the Higgs boson represents, in fact, a consequence
but not a cause of the spontaneous breaking, i.e. effective hiding
of the gauge symmetry. As it has been shown by t'Hooft [6] only
exactly unbroken gauge symmetry admits re-normalization of the
electro-weak interaction. Vice versa, any dynamical breaking of
the gauge symmetry causes a principal non-re-normalizability of
the electro-weak interaction. Similarly, comparison of the
spontaneous gauge symmetry breaking with  really dynamical
symmetry breaking by simple classical mechanical systems, eg. a
pellet in the bottle with a rotating double well bottom, has only
a limited validity. Spontaneous breaking of the gauge symmetry is
really caused, on the one hand, by local divergence of the small
perturbation approximation of the quantum field dynamics within a
small vicinity of the field corresponding to false vacuum. On the
other hand it is caused by local convergence of the same
approximate dynamics within a small vicinity of the field
corresponding to degenerate minimum of the potential energy
density. Or, spontaneous breaking or effective hidding of the
gauge symmetry exists only within small perturbation approximation
of the quantum field dynamics. It can be added and pointed out
that exact solution of given dynamics cannot be practically
obtained.) It is not hard to see that spontaneous breaking
(effective hiding) of the dynamical symmetries represents a
consistent physical model of the mathematical theory of the
probability (statistics) [4].

In this work an especial but very interesting case of the general
formalism of the spontaneous symmetry breaking (effective hiding)
will be considered. Concretely, spontaneous (non-dynamical)
breaking (effective hiding) of the unitary quantum mechanical
dynamical symmetry, precisely superposition of the weakly
interfering wave packets will be considered. Conceptual analogies
with spontaneous breaking of the gauge symmetry in
Weinberg-Sallam's electro-weak interaction are pointed out. Also,
consequences of the spontaneous superposition breaking in the
measurement process will be discussed. Namely, it is well-known
[7],[8] the following. Any dynamical breaking (by some hidden
dynamical terms, so-called hidden variables) of the unitary
quantum mechanical dynamical symmetry (superposition), i.e.
collapse in the quantum measurement must be either inconsistent
(contradictory to experimental data) or physically implausible
(superluminal). It will be demonstrated that quantum measurement
can be consistently and physically plausible considered as a
typical Landau's continuous phase transition. Or, collapse by
measurement can be considered  as the spontaneous breaking of the
unitary quantum mechanical dynamical symmetry (superposition).
Here critical values of the order parameters are determined by
Heisenberg's uncertainty relations.

\section {Spontaneous superposition breaking on a simple quantum mechanical system}

As it is well-known within standard quantum mechanical formalism
[12]-[15] Schr$\ddot{o}$dinger's equation
\begin {equation}
     \hat {H}\Psi \equiv (\frac {\hat {p}^{2}}{2m} + \hat {V}) \Psi = \hat {E}\Psi
\end {equation}
represents the exact (non-relativistic) dynamical equation of a
simple (without subsystems) quantum system. Here $\hat {H} =
(\frac {\hat {p}^{2}}{2m} + \hat {V}$ represents the Hamiltonian
observable of given system, $\hat {p} = -i \hbar \frac
{\partial}{\partial x}$  - momentum observable of given system,
$\hat {V}= \hat {V}(x)$ - potential energy observable of given
system, $\hat = i \hbar \frac {\partial}{\partial t}$ - total
energy observable of given system, $m$ - mass of given system,
$\hat {x}= x$ - coordinate observable of given system, $t$ - time
moment,  $\hbar$ - reduced Planck's constant, and, $\Psi = \Psi
(x,t)$ - exact quantum mechanical dynamical state of given system.
This state belongs to  Hilbert's space ${\it H}$ and satisfies
unit norm condition
\begin {equation}
     \int |\Psi|^{2} dx = 1
\end {equation}
(For reason of the simplicity, without any diminishing of the
generality of the basic conclusions, only one-dimensional, along
$x$ axis classically speaking, dynamics of the conservative (with
time independent $\hat {V}$) quantum system will be considered.)
General solution of (1), in common with initial condition
\begin {equation}
     \Psi(x,0) = \Psi_{0}
\end {equation}
, determines unambiguously (neglecting constant phase factor) the
quantum mechanical dynamical state in any time moment. It,
formally, can be presented by
\begin {equation}
     \Psi(x,t) \equiv \hat {U}\Psi = \exp[\frac {\hat {H}t}{i \hbar}] \Psi_{0}
\end {equation}
where $\hat {U} = \exp[\frac {\hat {H}t}{i \hbar}]$ represents the
unitary quantum mechanical dynamical evolution operator.

Chose a time dependent  basis ${\it B}(t) = { \Psi_{n} \equiv
\Psi_{n} (x,t) = \hat {U}\Psi_{n}(x,0) , \forall n}$ in ${\it H}$.
Obviously, given basis dynamically evolves analogously to (4). It
implies that (1) can be equivalently transformed in
\begin {equation}
\sum_{n} (\Psi_{n},\Psi) (\frac {\hat p^{2}}{2m} + \hat {V})
\Psi_{n} =  \sum_{n} (\Psi_{n},\Psi) \hat {E}\Psi_{n}
\end {equation}
Here superposition coefficients $(\Psi_{n},\Psi)$ for $\forall n$
satisfy the unit norm condition corresponding to (2)
\begin {equation}
\sum_{n}|(\Psi_{n},\Psi)|^{2} = 1
\end {equation}
as well as condition of the time independence
\begin {equation}
(\Psi_{n}(x,t),\Psi (x,t)) = (\hat {U}\Psi_{n}(x,0), \hat {U}\Psi
(x,0)) = (\Psi_{n}(x,0),\Psi(x,0)) = const \hspace{1cm} for
\forall n
\end {equation}

Obviously ${\it B}(0)$ can be chosen quite arbitrarily. It
implies, practically,  that exact unitary quantum mechanical
dynamics conserves the superposition of the states from  any
complete basis in ${\it H}$. Or, more generally, it means that all
bases in ${\it H}$  represent the equivalent referential frames
for description of the exact quantum mechanical dynamics. Also,
between two bases in ${\it H}$  there is one-to-one mapping by
corresponding unitary transformation. It means that invariance of
the exact quantum mechanical dynamics in respect to arbitrary
unitary transformation (representing, roughly speaking, a
"rotation" in the Hilbert's space), called unitary symmetry,
represents a basic exact dynamical symmetry of the exact quantum
mechanical dynamics.

It can be observed that, in some degree, interference terms in
(8), (9) are similar to Goldstone modes in the physical theories
with spontaneous symmetry breaking [1]-[3]. Namely, as it has been
discussed previously, interference terms point out implicitly on
the existence of the unitary symmetry of the quantum mechanical
dynamics. Given symmetry represents, roughly speaking, a symmetry
in respect to "rotations" in Hilbert space. Similarly, Goldstone
modes point out implicitly on the rotational symmetry in different
physical theories with spontaneous symmetry breaking. Also, choice
of the eigen basis of given observable within which interference
terms disappear is, in some degree, similar to Higgs mechanism. It
represents the choice of an especial rotating referential frame
within which Goldstone modes disappear. But existence of a more
detailed conceptual similarity between quantum mechanics and
physical theories with spontaneous symmetry breaking needs an
additional analysis. Especially, it needs a discussion on the
possibility of the spontaneous breaking of the unitary symmetry,
i.e. superposition within some approximation of the quantum
mechanical dynamics.

Consider, now, quantum mechanical dynamical state $\Psi$ from
orbital Hilbert's space ${\it H}_{ORB}$. Orbital Hilbert's space
${\it H}_{ORB}$  represents an especial sub-space of the usual
Hilbert's space ${\it H}$ . Over this sub-space only such
observable $\hat {A}$ acts that represents an analytical function
of the coordinate or/and momentum observable. As it is well-known,
average value and standard deviation of $\hat {A}$  in $\Psi$ are
given by expressions
\begin {equation}
<\hat {A}> = (\Psi , \hat {A}\Psi) =  \int \Psi * \hat {A}\Psi dx
\end {equation}
\begin {equation}
\Delta \hat {A} = (<\hat {A}^{2}> - <\hat {A}>^{2})^{\frac {1}{2}}
\end {equation}
Average value of an observable, including $\hat {A}$, in $\Psi$
has numerically identical values in all bases. But this average
value can be consistently interpreted as the statistical average
value only in the eigen basis of given observable. In this basis
interference terms, i.e. nondiagonal elements do not exist. In all
other bases, where corresponding interference terms exist, given
average value cannot be interpreted consistently statistically. It
refers on the expressions (8),(9) too.

According to (8),  Schr$\ddot {o}$dinger's exact quantum
mechanical dynamics for quantum state (1),(3) can be changed by
equivalent Ehrenfest's exact quantum mechanical dynamics for
average values
\begin {equation}
    m \frac {d<x>}{dt} = <\hat {p}>
\end {equation}
\begin {equation}
\frac {d<\hat {p}>}{dt} = - < \frac {\partial \hat
{V}(x)}{\partial x}>
\end {equation}
\begin {equation}
(\Psi (x,0),x \Psi (x,0)) = (\Psi _{0},x \Psi_{0}) \equiv x_{0}
\end {equation}
\begin {equation}
(\Psi(x,0),\hat {p}\Psi(x,0)) = (\Psi_{0}, \hat {p}\Psi_{0})
\equiv p_{0}
\end {equation}
Here (10,(13) represent corresponding initial conditions.

Exact solution of the system (10)-(13) considers general solution
of (10),(11) and introduction of the initial conditions (12),(13)
in this general solution. But this exact solution is, mostly,
mathematically very complicated. For this reason only some simple
approximate solutions of (10)-(13) are appropriate for
consideration. For example it can be supposed that $\hat {A}$
represents a function of $x$ only (dependence of $\hat {p}$ can be
formally neglected without any diminishing of the generality of
the basic  conclusions). Then (8) can be  Taylor expanded in a
relatively small vicinity of $<x>$, i.e. within the following
coordinate interval
\begin {equation}
  X(\Psi) = ( <x> - \frac {\Delta x}{2} ,<x> + \frac {\Delta x}{2})
\end {equation}
It yields

\[<\hat {A}> = A(<x>) + [\frac {\partial A(<x>)}{
\partial<x>}]<x-<x>> + \frac {1}{2}[\frac {\partial
^{2}A(<x>)}{\partial <x>^{2}}]<x-<x>>^{2} + \cdot \cdot \cdot =\]
\begin {equation}
 = A(<x>)  + \frac {1}{2} [\frac {\partial ^{2}A(<x>)}{\partial <x>^{2}}] \Delta x^{2} + \cdot \cdot \cdot
\end {equation}
In general case series (15) diverges. But series (15) can converge
for some exact quantum mechanical dynamical state $\Psi$ during
some finite but relatively large time interval  $T$. Especially
for some exact quantum mechanical dynamical state $\Psi$, during
$T$, for any observable $\hat {A}$ that acts over ${\it H}_{ORB}$
it can be satisfied
\begin {equation}
<\hat {A}> \simeq A(<x>)
\end {equation}
or, equivalently,
\begin {equation}
<\hat {A}^{2}> \simeq <\hat {A}>^{2}
\end {equation}
and
\begin {equation}
 |<\hat {A}>| \gg \Delta \hat {A}
\end {equation}
Given approximation conditions (16)-(18) define well-known wave
packet approximation [13],[14]. Exact quantum mechanical dynamical
state $\Psi$ that satisfies (16)-(18) represents approximately a
wave packet. Given wave packet, i.e. $\Psi$ in the wave packet
approximation will be denoted $\Psi _{WP}$. For a wave packet (see
for example detailed analysis in [13],[14]) Heisenberg's
uncertainty relation obtains the following form
\begin {equation}
\Delta x \Delta \hat {p}\simeq \hbar
\end {equation}
It means that wave packet is practically certainly placed in
$X(\Psi)$ so that (8) can be approximated by
\begin {equation}
<\hat {A}> \simeq (\Psi_{WP}, \hat {A} \Psi_{WP}) \simeq
\int_{X(\Psi)} \Psi _{WP}* \hat {A}\Psi _{WP} dx \simeq A(<x>)
|\Psi _{WP} |^{2}\Delta x \simeq A(<x>)
\end {equation}
Additionally it can be  supposed that $<\hat {A}>$ is
significantly time dependent while $\Delta \hat {A}$  is
practically time independent so that wave packet dissipates
extremely slowly. Then continuously time dependent $<\hat {A}>$
can be considered as an order parameter. Simultaneously $\Delta
\hat {A}$  can be considered as a practically constant critical
value of the order parameter. When absolute value of given order
parameter $<\hat {A}>$ is larger than its critical value $\Delta
\hat {A}$, i.e. when (16)-(18) are satisfied, approximate wave
packet dynamics  is self-complete. But, in the opposite case, when
absolute value of given order parameter $<\hat {A}>$ is smaller
than its critical value $\Delta \hat {A}$, i.e. when (16)-(18) are
unsatisfied, approximate wave packet dynamics becomes
self-incomplete. It admits a typical Landau's continuous phase
transition, by continuously time decreasing absolute value of
$<\hat {A}>$, from self-complete approximate wave packet dynamics
in the self-incomplete approximate wave-packet dynamics. And vice
versa.

In this way  within wave packet approximation equations
(10),(12),(13) stand practically unchanged while equation (11)
turns in the following equation
\begin {equation}
   \frac {d<\hat {p}>}{dt} = -  \frac {\partial V(<x>)}{\partial <x>}
\end {equation}
Simultaneously approximate dynamics determined by
(10),(12),(13),(21) becomes an effective dynamics. It corresponds,
obviously, to Newton's classical mechanical dynamics for a
particle with coordinate $<x>$ and  momentum $<\hat {p}>$ for
classical mechanical force $ -  \frac {\partial V(<x>)}{\partial
<x>}$.

It can be observed that within wave packet approximation unitary
symmetry of the exact quantum mechanical dynamics is not broken.
But it becomes implicit since average value of a quantum
mechanical observable does not depend of the choice of the basis
in the (orbital) Hilbert's space.

Thus, exact quantum mechanical dynamics is always  exactly
globally stable in  the orbital Hilbert's space. Also, within wave
packet approximation,  quantum mechanical dynamics can be
presented effectively by self-complete classical mechanical
dynamics globally stable within usual coordinate space. Vice
versa, when wave packet approximation is unsatisfied  exact
quantum mechanical dynamics cannot be approximated classical
mechanically dynamically globally stable in the usual coordinate
space. Finally, transition from the situation when wave packet
approximation is satisfied in the discretely different situation
when wave packet approximation is unsatisfied, and vice versa, can
be presented as a typical Landau's continuous phase transition. In
this transition order parameters have critical values determined
by Heisenberg's uncertainty relation. From effective classical
mechanical view point $|\hat {A}|$ represents the statistically
perturbed, i.e. with some absolute error proportional to $\Delta
\hat {A}$, estimated value of a classical variable $A$. It can be
additionally supposed that $|\hat {A}|$, in distinction from
$\Delta \hat {A}$, depends intensively of time and quantum
mechanical dynamical state $\Psi$. For this reason satisfaction of
(18) in the better and better approximation can be interpreted in
sense that absolute error of the statistical estimation of the
"exact" value of the classical mechanical variable $A$ is smaller
and smaller. And vice versa.

Further, suppose that there are two quantum states $\Psi_{1}$ and
$\Psi_{2}$ from ${\it H}_{ORB}$ that both satisfy wave packet
approximation. Then wave packets corresponding to these quantum
states, $\Psi_{1WP}$ and $\Psi_{2WP}$, are weakly interfering,
i.e. satisfy additional weak interference approximation, if the
following approximation condition is satisfied
\begin {equation}
   X(\Psi_{1}) \cap X(\Psi_{2}) = \emptyset
\end {equation}
where $\cap$ represents cross of the intervals and $\emptyset$-
empty interval. This condition can be presented in the equivalent
form
\begin {equation}
   |<x>_{1} - <x>_{2}| \geq \frac {\Delta_{1}x+\Delta_{2}x}{2}
\end {equation}
where
\begin {equation}
   <x>_{n} = (\Psi_{n}, x\Psi_{n})       \hspace{0.5cm}         for  \hspace{0.5cm} n = 1,2
\end {equation}
\begin {equation}
   \Delta_{n}x = (<x^{2}>_{n} - <x>^{2}_{n})^{1/2}   \hspace{0.5cm}  for \hspace{0.5cm} n = 1,2
\end {equation}
In further approximation
\begin {equation}
   \Delta_{n}x \simeq \Delta x \simeq const       \hspace{0.5cm}         for \hspace{0.5cm} n = 1,2
\end {equation}
(23) turns in
\begin {equation}
   |<x>_{1} - <x>_{2}| \geq \Delta x
\end {equation}
Obviously, two weakly interfering wave packets are classical
mechanically distinguishable.

Now, suppose that ${\it B}(t) = { \Psi_{n}\equiv \Psi_{n}(x,t) =
\hat {U}\Psi_{n}(x,0) , \forall n}$ represents a basis in ${\it
H}_{ORB}$ whose all quantum states satisfy wave packet
approximation. Suppose too  that wave packets $\Psi_{nWP}$  for
$\forall n$ corresponding  to quantum states from ${\it B}(t)$ are
all mutually weakly interfering. It means that the following
approximation conditions are satisfied
\begin {equation}
   (\Psi_{n}, \hat {A}\Psi_{m}) \simeq (\Psi_{nWP},\hat {A}\Psi_{mWP}) \simeq A(<x>_{n}) \delta_{nm}  \hspace{0.5cm}       for \hspace{0.5cm} \forall n,m
\end {equation}
\begin {equation}
   |<x>_{n} - <x>_{m}| \geq \Delta x \simeq \Delta_{n}x \simeq \Delta_{m}x  \hspace{0.5cm} for \hspace{0.5cm} \forall n,m
\end {equation}
where
\begin {equation}
    <\hat {D}>_{n} = (\Psi_{n},\hat {D}\Psi_{n}) \hspace{0.5cm} for \hspace{0.5cm} \forall n
\end {equation}
and
\begin {equation}
     \Delta_{n}\hat {D} = (<\hat {D}^{2}>_{n} - <\hat {D}>^{2}_{n})^{1/2} \hspace{0.5cm} for  \hspace{0.5cm} \forall n
\end {equation}
for arbitrary observable $\hat {D}$, eg. $x$, that acts over ${\it
H}_{ORB}$. Here $\delta_{nm}$ represents well-known Kronecker's
symbol. Then exact quantum mechanical dynamical state $\Psi = \Psi
(x,t)$   can be exactly quantum mechanically presented as the
following superposition of the quantum states from ${\it B}(t)$
\begin {equation}
   \Psi = \sum_{n} (\Psi_{n},\Psi) \Psi_{n} = \sum_{n}c_{n}\Psi_{n}
\end {equation}
Here
\begin {equation}
   c_{n} = (\Psi_{n},\Psi)  \hspace{0.5cm}      for  \hspace{0.5cm} \forall n
\end {equation}
represent superposition coefficients. It will be supposed that
$\Psi$ does not belong to ${\it B}(t)$ so that superposition (32)
is nontrivial. Or, it will be supposed that at least two
superposition coefficients in (32) are different from zero. Then
it can be simply proved that $\Psi$, i.e. a nontrivial
superposition of the wave packets, does not represent any wave
packet (does not satisfy the wave packet approximation). Namely,
according to suppositions, it follows
\begin {equation}
   <\hat {A}^{2}> \simeq \sum_{n} |c_{n}|^{2} A^{2}(<x>_{n})
\end {equation}
\begin {equation}
   <\hat {A}>^{2}\simeq \sum_{n} |c_{n}|^{2}A(<x>_{n}) <\hat {A}>
\end {equation}
Here, in general case, superposition coefficients and theirs
absolute values must be independent of  $A(<x>_{n})$  for $\forall
n$. Since $\hat {A}$ can be arbitrary observable that acts over
${\it H}_{ORB}$  it can be chosen that
\begin {equation}
   A(<x>_{n}) > 0    \hspace{0.5cm}     for \hspace{0.5cm} \forall  n
\end {equation}
But then wave packet approximation condition (17) can be satisfied
only for an especial condition
\begin {equation}
   A(<x>_{n}) \simeq <\hat {A}>  \hspace{0.5cm}  for  \hspace{0.5cm} \forall  n
\end {equation}
I.e. (17) cannot be satisfied for all observable that act over
${\it H}_{ORB}$ and satisfy (36). It, simply, means that $\Psi$
cannot satisfy wave packet approximation. Or, it means that
$\Psi$, that exactly represents a quantum mechanically dynamically
globally stable state, represents a classical mechanically
dynamically globally unstable state.

Nevertheless $\Psi$  represents the superposition of the weakly
interfering, i.e., classically speaking, separated in the
coordinate space, wave packets. For this reason $\Psi$ can be
considered as classical mechanically dynamically locally stable in
a small vicinity of any wave packet. Precisely, $\Psi$ is
classical mechanically dynamically stable within coordinate
intervals
\begin {equation}
  X_{red}(\Psi_{n}) = ( <x>_{n} - \frac {\Delta_{n red}x}{2} , <x>_{n} + \frac {\Delta_{n red}x}{2})   \hspace{0.5cm} for \hspace{0.5cm} \forall  n               .
\end {equation}
representing reductions of the coordinate intervals $X(\Psi_{n})$
for $\forall n$. Obviously, $ X_{red}(\Psi_{n})$ and $X(\Psi_{n})$
have the same centrum $<x>_{n}$  (which implies that within both
intervals the same wave packet dynamics, that explicitly depends
of the average value of the coordinate, is satisfied) for $\forall
n$. But it will be supposed that width of $X_{n red}(\Psi_{n})$ ,
$\Delta_{n red}x$, is smaller than width of $X(\Psi_{n})$,
$\Delta_{n}x$, i.e. that it represents a reduction of
$\Delta_{n}x$  , for $\forall n$ . Then, according to wave packet
approximation, it follows
\begin {equation}
   <x>_{n} \gg \Delta_{n}x > \Delta_{n red}x   \hspace{0.5cm}   for  \hspace{0.5cm} \forall n
\end {equation}
which means that supposed local reductions of the wave packets
widths are unobservable for wave packet dynamics. Simply speaking,
in difference from exact quantum mechanical dynamics wave packet
dynamics cannot differ wave packet and reduced wave packet (wave
packet in corresponding reduced coordinate interval).

It will be supposed that by given local reductions the following
condition of the local conservation of the average value is
satisfied
\begin {equation}
   |c_{n}|^{2} A(<x>_{n}) \Delta_{n}x  =   A(<x>_{n}) \Delta_{n red}x   \hspace{0.5cm}    for \hspace{0.5cm} \forall n
\end {equation}
It, on the one side, determines widths of the reduced coordinate
intervals according to suppositions
\begin {equation}
   \Delta_{n red}x  =  |c_{n}|^{2} \Delta_{n}x  <  \Delta_{n}x        \hspace{0.5cm}   for \hspace{0.5cm} \forall  n
\end {equation}
On the other hand it yields
\begin {equation}
    |c_{n}|^{2} = \frac {\Delta_{n red}x }{\Delta_{n}x }   \hspace{0.5cm} for \hspace{0.5cm} \forall n
\end {equation}
It represents a very interesting result. Namely $\Delta_{n red}x $
represents  the measure of  $ X_{red}(\Psi_{n})$ while
$\Delta_{n}x $ represents the measure of $ X(\Psi_{n})$ for
$\forall n$. It admits, in common with normalization condition
\begin {equation}
    \sum_{n} |c_{n}|^{2} = 1              ,
\end {equation}
that  $|c_{n}|^{2}$ for $\forall n$ be consistently considered as
"geometrically" defined probabilities [5].

Now, the meaning of the events corresponding to given
probabilities will be explained. Firstly, it can be pointed out
that given superposition of the weakly interfering wave packets
cannot be simply split in the simultaneously existing wave packets
with corresponding intensities. Namely, splitting of the weakly
interfering wave packets in the simultaneously existing wave
packets with corresponding intensities smaller than 1 implies
unambiguously exact breaking of the unitary symmetry of the
quantum mechanical dynamics. More precisely, such splitting breaks
the unit norm of the quantum dynamical state. So, conservation of
the unitary symmetry of the exact quantum mechanical dynamics
forbids definitely splitting of the superposition of the weakly
interfering wave packets in the simulatenous existing wave packets
with corresponding intensities, i.e. norms smaller than 1. In this
way mentioned probabilistic events cannot be explained by
splitting of the superposition of the weakly interfering wave
packets in the simultaneously existing wave packets.

Nevertheless, there is a possibility that superposition of the
weakly interfering wave packets is represented by arbitrary wave
packet. Such really local representation, i.e. projection, can be
approximately, i.e. effectively presented as a global
representation by means of characteristic probabilistic, i.e.
statistical methods. Precisely, as it has been pointed out
previously, in difference from exact quantum mechanical dynamics
wave packet dynamics does not differ wave packet and its local
reduction. For this reason, from aspect of the wave packet
dynamics, change of the globally unstable superposition of the
weakly interfering wave packet by a locally stable reduced wave
packet can be effectively presented as change of given
superposition by corresponding globally stable wave packet.
Moreover, it is not hard to see that quotient of the coordinate
interval of a wave packet and coordinate interval corresponding to
superposition (representing, in fact, union of the coordinate
intervals of all wave packets) is equivalent to quotient of the
coordinate interval of corresponding reduced wave packet and
coordinate interval of given wave packet (42). This quotient, of
course, represents probability of the event that from aspect of
the wave packet dynamics globally unstable superposition of the
wave packets is changed by one locally stable reduced wave packet,
i.e. effectively by corresponding globally stable wave packet.

It is very important to be pointed out that given events exist,
i.e. can occur, only effectively, within wave packet
approximation. Appearance of given events within given
approximation is called self-collapse.Without given approximation,
i.e. completely exactly, quantum mechanically, self-collapse does
not exit at all. For this reason self-collapse does not any
influence (change) on the exactly existing globally stable quantum
mechanical dynamical state $\Psi$ (32).

Thus, superposition of the weakly interfering wave packets and
superposition breaking, i.e. self-collapse, exist simultaneously
but on the discretely different levels of the analysis accuracy,
exact and approximate. It represents a typical "relativistic"
effect in sense that the dynamical concepts depend significantly
from the level of the analysis accuracy [16],[17]. Similar
situation exists in the general theory of relativity. Here,
exactly speaking, there is no gravitational force in the absolute
Euclidian space and time, but there is an Einstein's gravitational
field in corresponding curved Riemanian space-time. Nevertheless,
in the approximation of the weak gravitational field Newton's
gravitational force in the absolute Euclidian space and time
effectively appears.

It is not hard to see that self-collapse can be considered as the
spontaneous (non-dynamical) unitary symmetry, i.e. superposition
breaking (effective hiding). In this sense self-collapse
represents an especial case of the general theory of the
spontaneous, i.e. nondynamical symmetry braking (effective hiding)
[1]-[3].

More precisely, here expression (43) determines the a-priory
probabilities of potential events. Given probabilities, for a
nontrivial superposition, are smaller than 1. But, by spontaneous
realization of the events, a-priory probabilities turn in the a
posteriori probabilities. Then a-posteriori probability of the
spontaneously realized event equals 1 while a-posteriori
probabilities of the unrealized events equal 0. In this way unit
norm of the quantum dynamical state is conserved. I.e. norm of the
superposition of the weakly interfering wave packets as well as
norms of the spontaneously realized wave packet equal 1. (Now it
can be observed that condition on the unit norm of the quantum
dynamical state represents a more precise form of the early
expression "indivisibility of the quant of action" and similar
[17]. Of course, this condition represents more precise form of
the atomistic concepts in the antique.)

However, it seems that "final" wave packet is principally
different from "initial" superposition of the weakly interfering
wave packets. For this reason it seems that here "initial"
superposition is "finally" broken as well as unitary symmetry.
Meanwhile such breaking is non-dynamical, i.e. spontaneous
(accidental) in full agreement with general theory of the
spontaneous symmetry breaking [1]-[3]. Or, here exact quantum
mechanical superposition is not exactly broken. It is only
effectively hidden in spontaneously realized wave packet by
statistical methods within wave packet approximation. Without
given approximation, i.e. exactly quantum mechanically
dynamically, superposition of the weakly interfering wave packets
is completely conserved. In other words, quantum system is always
exactly quantum mechanically dynamically described by
superposition of the weakly interfering wave packets.
Simultaneously, but within wave packet approximation, it is
described by one, spontaneously realized, wave packet. For this
reason superposition of the weakly interfering wave packets and
spontaneously realized wave packet do not belong to the same level
of the analysis accuracy. It implies that spontaneously realized
wave packet represents the most appropriate representation, i.e.
projection of the exact superposition of the weakly interfering
wave packets at approximate level of the analysis accuracy. Or,
all wave packets represent different representations, i.e.
projections of the same exact superposition of the weakly
interfering wave packets at the approximate level of the analysis
accuracy. In other words, unique exact superposition of the weakly
interfering wave packets corresponds to a degenerate set of its
projections, i.e. wave packets at the approximate level of the
analysis accuracy.

Finally, it is very important to be pointed out that basis of the
weakly interfering wave packets represents unique basis over which
superposition breaking occurs within wave packet approximation.
Namely, any other basis, principally different from given basis,
holds quantum mechanical states that represent nontrivial
superposition of the weakly interfering wave packets. For this
reason, within wave packet approximation, any quantum mechanical
state from any other basis, principally different from given
basis, turns spontaneously (non-dynamically and probabilistically)
in a wave packet from basis of the weakly interfering wave
packets.

\section {Spontaneous superposition breaking on a quantum mechanical super-system. Measurement as a Landau's continuous phase transition }

Suppose that there is a quantum mechanical super-system, $1+2$,
that holds two quantum mechanical sub-systems, $1$ and $2$.
Suppose that quantum mechanical dynamical state of $1+2$
represents the following non-trivial correlated (entangled)
quantum state (non-trivial super-systemic superposition)
\begin {equation}
     \Psi^{1+2} =  \sum_{n} c_{n} \Psi^{1}_{n}\otimes \Psi^{2}_{n}
\end {equation}
from Hilbert's space of the quantum states of $1+2$, ${\it
H}_{1+2}= {\it H}_{1}\otimes {\it H}_{2}$. Here ${\it B}_{1} = {
\Psi^{1}_{n}, \forall n}$ represents a basis in the Hilbert's
space of the quantum states of $1$, ${\it H}_{1}$ , ${\it B}_{2} =
{ \Psi^{2}_{n}}, \forall  n$ - a basis in the Hilbert's space of
the quantum states of $2$ , ${\it H}_{2}$ , $\otimes$ - tensorial
product , $c_{n}$   for $\forall n$   - superposition coefficients
that satisfy unit norm condition
\begin {equation}
    \sum_{n} |c_{n}|^{2} = 1
\end {equation}
According to standard quantum mechanical formalism [12]-[15], its
usual interpretations [16],[17] as well as later theoretical [6]
and experimental analysis [7], $\Psi^{1+2}$ describes exactly
(completely) and dynamically stable $1+2$  inseparable in $1$ and
$2$.

Suppose that ${\it H}_{2}$   represents the orbital Hilbert's
space , ${\it H}_{2ORB}$  , and that ${\it B}_{2}$  represents the
basis of the weakly interfering wave packets so that conditions
analogous to (28),(29)
\begin {equation}
   (\Psi^{2}_{n}, \hat {A}_{2}\Psi^{2}_{m}) \simeq (\Psi^{2}_{nWP},  \hat {A}_{2}\Psi^{2}_{mWP}) \simeq A_{2}(<x_{2}>_{n}) \delta_{nm}
   \hspace{0.5cm}     {\rm for}  \hspace{0.5cm}\forall n,m
\end {equation}
\begin {equation}
   |<x_{2}>_{n} - <x_{2}>_{m}| \geq \Delta x_{2} \simeq \Delta_{n}x_{2} \geq \Delta_{m}x_{2}   \hspace{0.5cm} for \hspace{0.5cm} \forall n,m
\end {equation}
is satisfied, where $\hat {A}_{2}$ represents arbitrary
observable, eg. $x_{2}$ that acts over ${\it H}_{2ORB}$. Also,
suppose that ${\it H}_{1}$ can be any Hilbert's space and ${\it
B}_{1}$ any basis in ${\it H}_{1}$.

Then, as it is not hard to see, in a complete analogy with
previous discussion, it can be consistently concluded the
following. Within wave packet approximation, discretely different
from the exact quantum mechanical description, globally
dynamically unstable state corresponding to super-systemic
non-trivial superposition (43) can be changed spontaneously
(non-dynamically) by some non-correlated (non-entangled) state
$\Psi^{1}_{n}\otimes \Psi^{2}_{nWP}$ with probability
$|c_{n}|^{2}$ for $\forall n$. Also, within given approximation,
non-correlated state $\Psi^{1}_{n}\otimes \Psi^{2}_{nWP}$ admits a
separation of  $1+2$ in $1$ and $2$. By this separation $1$ is
described effectively exactly quantum mechanically (without any
numerical approximation of the quantum state) by $\Psi^{1}_{n}$
while $2$  is described effectively within wave packet
approximation by $\Psi^{2}_{nWP}$ for $\forall n$. In this way
here a sub-systemic spontaneous (non-dynamical) superposition
breaking, i.e. self-collapse on the one  sub-system (eg. $2$)
occurs if condition of the weakly interfering wave packet
approximation is satisfied sub-systemically. Simultaneously,
relatively, i.e. in respect  to self-collapsed quantum mechanical
sub-system (eg. $2$), other quantum mechanical sub-system (eg.
$1$) is described by an effectively exact (without any numerical
approximation) mixture of the quantum mechanical states from
correlated basis. Or, it can be stated that on this sub-system a
relative (in respect to the self-collapsed first sub-system)
superposition breaking. i.e. collapse appears. For this reason
given relative superposition breaking, i.e. relative collapse
represents an effective exact (without any numerical
approximation) quantum mechanical phenomenon on corresponding
subsystem (eg. $1$).

Now, according to some previous ideas [9]-[11], it will be proved
that sub-systemic spontaneous (non-dynamical) superposition
breaking, i.e. self-collapse and relative collapse on the
correlated quantum mechanical sub-systems, can be used for the
consequent and consistent description of the measurement process
[8],[12],[16],[17].

Suppose that $1$ represents the measured quantum object. Suppose
that, before measurement, $1$ is exactly quantum mechanically
described  by quantum mechanical dynamical state
\begin {equation}
  \Psi^{1} =  \sum_{n} c_{n}\Psi^{1}_{n}
\end {equation}
where ${\it B}_{1} = { \Psi^{1}_{n}, \forall n}$ represents a time
independent eigen basis of the measured observable. Also suppose
that $2$ represents the measuring apparatus. Suppose that, before
measurement, $2$  is exactly quantum mechanically described by
quantum mechanical dynamical state
\begin {equation}
  \Psi^{2} =  \Psi^{2}_{0}
\end {equation}
where ${it B}_{2} = { \Psi^{2}_{n} , \forall n}$ represents, on
the one hand, a time dependent eigen basis of so-called pointer
observable and, on the other hand, basis whose states represent
the wave packets. Then, before measurement, $1+2$ is exactly
quantum mechanically described by the following non-correlated
(non-entangled) quantum mechanical dynamical state
\begin {equation}
  \Psi^{1}\otimes \Psi^{2} =  \sum_{n} c_{n} \Psi^{1}_{n}\otimes \Psi^{2}_{0}           .
\end {equation}
Further, suppose that  exact quantum mechanical dynamical
interaction between $1$ and $2$ can be presented by von Neumann
unitary quantum mechanical dynamical evolution operator $\hat
{U}_{1+2}$ determined by
\begin {equation}
   \hat {U}_{1+2}\Psi^{1}_{n}\otimes \Psi^{2}_{m} = \Psi^{1}_{n} \otimes \Psi^{2}{m+n}  \hspace{0.5cm}       for \hspace{0.5cm}  \forall n,m                            .
\end {equation}
More precisely, it can be supposed that, during some relatively
small time interval $\tau$, $\hat {U}_{1+2}$ changes the initial
quantum mechanical state $\Psi^{1}_{n}\otimes \Psi^{2}_{m}$ in the
final quantum mechanical state  $\Psi^{1}_{n}\otimes
\Psi^{2}_{m+n}$  for $\forall n,m$. Then, after given interaction,
precisely  after time interval  $\tau$ or in time moment $\tau$,
$1+2$ is  exactly quantum mechanically described by the following
correlated (entangled) quantum mechanical dynamical state
\begin {equation}
     \Psi^{1+2} =  \sum_{n} c_{n} \Psi^{1}_{n}\otimes \Psi^{2}_{n}                 .
\end {equation}
Simply speaking by exact quantum mechanical dynamical interaction
between $1$ and $2$, i.e during time interval t,  an extension of
the superposition from $1$ on $1+2$ occurs.

Suppose that given interaction between $1$ and $2$ is chosen in
such way that after this interaction, i.e. after time moment
$\tau$, conditions of the weak interference between wave packets
from ${\it B}_{2}$, (46), (47), become satisfied. More precisely,
suppose that absolute values of the order parameters $<x_{2}>_{n}
- <x_{2}>_{m}$ for $\forall n, m$, according to appropriately
chosen exact quantum mechanical dynamics, i.e.  $\hat {U}_{1+2}$,
increase continuously during time. Suppose, further, that after
time moment $\tau$, their absolute values become larger than
critical value $\Delta x_{2}$ determined, implicitly, by
Heisenberg's uncertainty relations. Then within wave packet
approximation (52) can be changed  spontaneously (non-dynamically)
by some non-correlated (non-entangled) state $\Psi^{1}_{n}\otimes
\Psi^{2}_{nWP}$ with probability $|c_{n}|^{2}$  for $\forall n$.
In this way, within given approximation, during time interval
$\tau$, a typical Landau's continuous phase transition with
spontaneous (non-dynamical) superposition breaking (effective
hiding), precisely with self-collapse on $2$ and relative collapse
on $1$, occurs.

 It is not hard to see that given phase transition possesses all
 necessary characteristics of the real measurement process.
 Namely, $2$, i.e. measuring apparatus, is described effectively
 classical mechanically by a statistical mixture of the eigen states
 of the pointer observable only. Also,  $1$, i.e. measured quantum
 object,  is described effectively quantum mechanically by a statistical
 mixture of the eigen states of the measured observable only.
 Statistical weights, i.e. probabilities of the events, in both
 mixtures are numerically, i.e. quantitatively determined by superposition
 coefficients of the exact quantum mechanical dynamical state
 of $1$, i.e. measured quantum object before measurement (48)
 exclusively. But, qualitatively, statistical or probabilistic
 characteristics of corresponding superposition coefficients
 represent intermediate consequences of the self-collapse that
 appears on $2$, i.e.  measuring apparatus exclusively. For
 this reason spontaneous superposition breaking, i.e. self-collapse
 on $2$ and relative collapse on $1$, yield a consistent and
 consequent formalization of the measurement. Exact quantum
 mechanical dynamical interaction between measured quantum object
 and measuring apparatus on the one hand, and, measurement as
 the spontaneous superposition breaking on the other hand,
 exist simultaneously but on the discrete different levels
 of the analysis accuracy, exact and approximate.
 (It represents a consequent formalization and consistent
explanation of the Bohr's concept of the complementarity, i.e.
relative boundary between measured object and measuring apparatus
[16],[17].) Meanwhile, by usual measurements, where measuring
apparatus represents a macroscopic system, there are extreme
technical difficulties (eg. decoherence by interactions with
environment) for immediate experimental studying of the existing
dynamical interaction between  measured quantum object and
measuring apparatus, or, roughly speaking, mesoscopic and
macroscopic superposition. Nevertheless, there are many realizable
or realized experiments [18],[19] that affirm or would affirm
existence of the mesoscopic or macroscopic superposition. Or,
given experiments affirm or would affirm  less or more explicitly
existence of the correlated quantum mechanical dynamical state of
the quantum mechanical super-system (measured quantum object +
measuring apparatus). For example, in [18] is theoretically
discussed a potentially realizable experiment at extremely small
temperatures (necessary for elimination of the environmental
decoherence). Within given experiment there is a quantum
mechanical super-system that holds two sub-systems. First one
represents a photon that can be considered as $1$. Second one
represents a movable mesoscopic mirror of a typical Michelson's
interferometer that can be considered as $2$. Roughly speaking
exact quantum mechanical dynamical interaction between photon and
mirror periodically correlates (entangles) and decorrelates
(deentangles) given sub-systems. During a correlation period all
conditions for spontaneous superposition breaking, i.e.
self-collapse on the mirror and relative collapse (absence of the
superposition) on the photon, are satisfied. But during next
decorrelation period  a superposition on the photon appears. It
implies that in the previous correlation period super-system
photon+mirror has been realy in a correlated state, i.e. in the
state of a super-systemic superposition. (Detailed discussion of
[18] from aspect of the spontaneous superposition breaking concept
is given in [9].) In [19] a mesoscopic quantum mechanical
superposition on a super-system is realized without decoherence.

Generally speaking, concept of the spontaneous superposition
breaking  is able to solve all open problems of the foundations of
the quantum mechanics [8]. Especially, on the one hand, $\hat
{U}_{1+2}$ or correlations between ${\it B}_{1}$ and ${\it B}_{2}$
are exactly quantum mechanically uniquely determined (neglecting
constant phase factors). On the other hand, self-collapse on $2$
within localization of the weakly interfering wave packets
approximation appears over ${\it B}_{2}$ exclusively. For this
reason within given measurement any observable that does not
commute with measured observable or whose eigen basis is different
from ${\it B}_{1}$ cannot be measured at all. In other words  two
or many noncommuting observable cannot be simultaneously
statistically distributed. It implies that within standard quantum
mechanical formalism, including measurement as the spontaneous
superposition breaking, an inequality analogous to Bell' s
inequality for hidden variables [7] cannot be consistently
formulated (for details see [10]). Simply speaking quantum
mechanics represents a local theory, in full agreement with
existing experimental data [8]. (Usually used term "quantum
nonlocallity" refers, in fact, only on the hypothetical hidden
variables theories that should satisfy  existing experimental
data.)

\section {Conclusion}

In conclusion the following fundamental principles of the standard
quantum mechanical formalism can be shortly repeated and pointed
out. Hilbert's space represents exactly the fundamental physical
space. All bases in this space represent exactly the fundamental
referential frames (reference systems). Any referential frame can
be exactly transformed in some other by corresponding unitary
transformation (representing, roughly speaking, a "rotation" in
the Hilbert's space). Quantum mechanical system is exactly and
completely described by unitary symmetric (that conserves
superposition and does not prefer any referential frame as the
absolute) quantum mechanical dynamical state (of the unit norm).
This state and its dynamical evolution is determined unambiguously
(neglecting constant phase factor) by Schr$\ddot {o}$dinger's
equation (or equivalent equations) and initial condition only.
Physical characteristics of given quantum mechanical system in its
quantum mechanical dynamical state are exactly represented by
average values of corresponding Hermitian operators, observable.
Given characteristics evolve exactly deterministically during time
according to exact quantum mechanical dynamics, i.e. Ehrenfest's
equations. There is none other fundamental principle of the
standard quantum mechanics so that quantum mechanics represent an
objective, deterministic and local (without Bell's inequality
analog) physical theory.

Measurement, as it has been suggested in [16],[17], represents
only an effective phenomenon. It can be consequently modeled by
spontaneous (nondynamical) superposition breaking with inherent
statistical (probabilistic) characteristics. It considers the
self-collapse that appears exclusively over eigen basis of the
pointer observable on the effectively classical mechanical
described measuring device. Simultaneously it considers the
relative collapse that appears exclusively over eigen basis of the
measured observable at the effectively quantum mechanically
described measured quantum object. Thus preference of the eigen
basis of the measured obeservable occurs really by  measurement.
But given preference is implicitly formulated even by a demand
that average value of given observable in the quantum mechanical
dynamical state of the measured quantum object has form of a
statistical average value. Namely, numerical value of an
observable, including measured observable, is the completely same
in any basis. It expresses unitary symmetry of the quantum
mechanical dynamics. Meanwhile in a basis different from eigen
basis of the measured observable interference terms appear. They,
on the one hand, point out that unitary symmetry is exactly
conserved, i.e. that "rotation" from one in the other referential
frame is still possible. But, on the other hand, theirs presence
forbids that given average value be presented in the form of the
statistical average value in given basis different from eigen
basis of the measured observable. It is not hard to see that given
interference terms in the standard quantum mechanical formalism in
some degree are analogous to Goldstone modes in the general
formalism of the spontaneous symmetry breaking [1]-[3].

So, spontaneous (nondynamical) superposition breaking points out,
intermediately, that fundamental unitary symmetry of the quantum
mechanical dynamics cannot be exactly broken but only effectively
hidden in the approximate measurement process. Or, instead of one
(absolute) referential frame (basis of the measured observable by
collapse) there are all possible (relative) referential frames
(bases in the Hilbert's space) that describe equivalently the
unitary symmetric quantum mechanical dynamics. It is in full
agrement with the following Bohr's words: "Before concluding I
should still like to emphasize the bearing of the great lesson
derived from general relativity theory upon the question of
physical reality in the field of quantum theory. In fact,
notwithstanding all characteristic differences, the situation we
are concerned with in these generalizations of classical theory
presents striking analogies which have often been noted.
Especially, the singular position of measuring instrument in the
account of quantum phenomena, just discussed, appears closely
analogous to the well-known necessity in relativity theory of
upholding an ordinary description of all measuring processes,
including sharp distinction between space and time coordinates,
although very essence of this theory is the establishment of new
physical laws, in comprehension of which we must renounce the
customary separation of space and time ideas. The dependence of
the reference system, in relativity theory, of all readings of
scales and clocks may even be compared with essentially
uncontrollable exchange of the momentum or energy between the
objects of measurement and all instruments defining the space-time
system of the reference, which in quantum theory confront us with
the situation characterized by the notion of complementarity. In
fact this new feature of natural philosophy means a radical
revision of our attitude as regards physical reality, which may be
paralleled with the fundamental modification of all ideas
regarding the absolute character of physical phenomena, brought
about general theory of relativity." [16]

\section {References}

\begin {itemize}

\item [[1]]  J. Bernstein, Rev.Mod.Phys., {\bf 46}, (1974.), 7.
\item [[2]]  F. Halzen,  A. Martin, {\it Quarks and Leptons: An Introductory Course in Modern Particle Physics} (John Wiley and Sons, New York, 1987.)
\item [[3]]  L. H. Ryder, {\it Quantum Field Theory} (Cambridge University Press , Cambridge, 1987.)
\item [[4]]  L. D. Landau, E. M. Lifshitz, {\it Statistical Physics} (Pergamon Press, Oxford, 1960.)
\item [[5]]  M. Loeve, {\it  Probability Theory} (Springer-Verlag, New York, 1978.)
\item [[6]]  G. t'Hooft, Nucl. Phys., {\bf B35}, (1971.), 167.
\item [[7]]  J. S. Bell, Physics, {\bf 1}, (1964.), 195.
\item [[8]]  A. Aspect, P. Grangier, G. Roger, Phys.Rev.Lett., {\bf 47}, (1981.), 460.
\item [[9]]  V. Pankovi\'c, M. Predojevi\'c,  M. Krmar, {\it Quantum Superposition of a Mirror and Relative Decoherence
(as Spontaneous Superposition Breaking) } quant-ph/0312015.
\item [[10]]  V. Pankovi\'c, T. H$\ddot{u}$bsch, M. Predojevi\'c,  M . Krmar,
{\it From Quantum to Classical Dynamics : A Landau Phase Transition with Spontaneous Superposition Breaking}, quant-ph/0409010 .
\item [[11]]  V. Pankovi\'c, M. Predojevi\'c,  M. Krmar, Hadronic Journal Supplement, {\bf 18}, (2003.), 295.
\item [[12]]  J. von Neumann, {\it  Mathematische Grundlagen der Quanten Mechanik} (Spiringer Verlag, Berlin, 1932.)
\item [[13]]  P. A. M. Dirac, {\it Principles of Quantum Mechanics} (Clarendon Press, Oxford, 1958.)
\item [[14]]  A. Messiah, {\it  Quantum Mechanics} (North-Holand Publ. Co., Amsterdam, 1970.)
\item [[15]]  B. d'Espagnat, {\it Conceptual Foundations of the Quantum Mechanics} (Benjamin , London-Amsterdam-New York, 1976.)
\item [[16]]  N. Bohr, Phys.Rev., {\bf 48}, (1935.), 696.
\item [[17]]  N. Bohr, {\it Atomic Physics and Human Knowledge} (John Wiley, New York, 1958.)
\item [[18]]  W. Marshall, C. Simon, R. Penrose, D. Bouwmeester, Phys.Rev.Lett. , {\bf 91}, (2003.), 130401.
\item [[19]]  F. De Martini, F. Sciarrino, V. Secondi, {\it Realization  of a Decoherence-free, Optimally Distingushable Mesoscopic Quantum Superposition }, quant-ph/0508196.

\end {itemize}

\end {document}